\documentclass[aps,prl,twocolumn,groupedaddress]{revtex4}

\usepackage{graphicx}
\usepackage{textcomp}
\newcommand{\ket}[1]{|{#1}\rangle}

\begin{document}

%\preprint{}

\title{Excitation of a single atom with exponentially rising light pulses}

\author{Syed Abdullah Aljunid, Gleb Maslennikov, Yimin Wang}
\altaffiliation{Center for Quantum Technologies, 3 Science Drive 2, Singapore, 117543}
\author{Dao Hoang Lan}
\altaffiliation{University of Twente}
\author{Valerio Scarani}
\author{Christian Kurtsiefer}
\altaffiliation{Center for Quantum Technologies and Department of Physics,
  National University of Singapore, 3 Science Drive 2,  Singapore, 117543}
\email[]{christian.kurtsiefer@gmail.com}
\date{\today}

\begin{abstract}
We investigate the interaction between a single atom and optical pulses in a
coherent state with a controlled temporal envelope. In a comparison between a
rising exponential and a square envelope,
we show that the rising exponential envelope leads to a higher excitation
probability for fixed low average photon numbers, in accordance to a time-reversed
Weisskopf-Wigner model.
We characterize the atomic transition dynamics for a wide range of the
average photon numbers, and are able to saturate the optical transition of a
single atom with $\approx$50 photons in a pulse by a strong focusing
technique. For photon numbers of $\approx$1000 in a 15\,ns
long pulse, we clearly observe Rabi oscillations.
\end{abstract}

\pacs{37.10.Gh, % Atom traps and guides
42.50.Ct,       % Quantum description of interaction of light and matter;
                % related experiments
32.90.+a        % Other topics in atomic properties and interactions of atoms;
                % with photons (restricted to new topics in section 32)
}

\maketitle

\section{Introduction\label{intro}}

In order to compose more and more complex networks of quantum systems for
quantum information processing, efficient interfaces between
different physical systems are required~\cite{cirac:97, kimble:08, dlcz:01}.
An important representative of such an interface are two-level atoms coupling
to photons that can propagate between distant atoms.

The fundamental processes for exchanging information between atoms and photons
are emission and absorption. While capturing an emitted photon from an atom
can usually be done with a high efficiency, the reverse process is more
challenging, since the field strength of a single photon is very
weak. Accomplishing a high excitation probability for an atom from a single
photon is thus quite challenging.
It is common to solve this problem in a context of cavity QED, where the field strength
of single photons at the location of the atom is dramatically increased by using optical cavities
with small mode volumes~\cite{Kimble:98}. However, sophisticated highly reflective
dielectric coatings are required to decouple the cavity from environmental losses which compromise
the scaling of such systems. To relax the coating requirements, the mode
volume has to be further decreased, and several experimental efforts target this issue~\cite{hunger:10}.
Placing an atomic 2-level system in a strongly focused mode also increases the
electrical field of a photon, and can lead to reasonably strong
interaction~\cite{tey:08, wrigge:08, Vamivakas:07} even without an optical
cavity. In this case, the
emission and absorption of photons are not affected by presence of artificial
boundary conditions, and the absorption only depends
on the {\it overlap} of spatial and frequency modes of the light field with
atomic transition modes.
Considering only dipole-allowed transitions and a lifetime-limited spectral
absorption profile, it has been shown that near perfect excitation probability
can be achieved with a wave packet that has an exponentially rising temporal envelope with a characteristic time on the order of the decay time of
the excited atomic state~\cite{Stobinska:09, Wang:11}. 

In this letter we investigate the effect of temporal shaping of light pulses on the excitation probability of a closed
cycling 2-level transition in a single $^{87}$Rb atom.

% \section{Experimental Setup\label{setup}}
\section{Experiment}

\begin{figure}
  \begin{center}
    % version v3 moved the point A to the atom, v2 has it outside
    % \includegraphics[width=0.9\columnwidth]{./figures/setupv3.eps}
    \includegraphics[width=0.9\columnwidth]{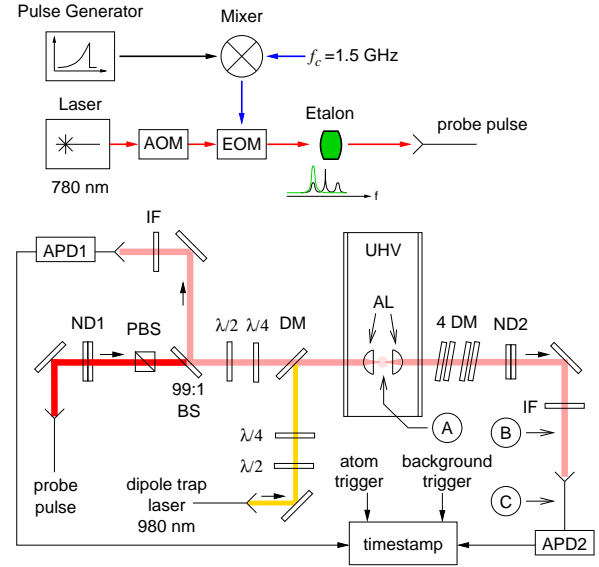}
  \end{center}
  \caption{Top: Preparation of pulses with controllable waveform. Bottom:
    Setup for transmission and reflection measurement of light by a single
    atom. UHV: ultra high vacuum chamber, AL: aspheric lenses with full
    NA=0.55 and focal length $f=4.51$\,mm, PBS: polarizing beam splitter, ND
    1,2: stacks of neutral density filters, $\lambda/2, \lambda/4$: wave
    plates, DM: dichroic mirrors, IF: interference bandpass filters centered
    at 780\,nm.
    \label{exp_setup}}
\end{figure}

The experimental setup is schematically shown in Figure~\ref{exp_setup}. A
single atom is trapped in a far-off resonant optical dipole trap (FORT) at the focus of two confocally
positioned aspheric lenses.
The FORT is loaded from a magneto-optical trap (MOT) holding $\approx 10^4$
atoms. A collisional blockade mechanism ensures that either zero or
one atom is trapped at any instance~\cite{schlosser:01}. A probe
beam is delivered from a single mode optical fiber and defines the focused
light mode that is coupled to the atom. This gives the probe beam a Gaussian
spatial mode with a characteristic waist $w_L=1$\,mm at the focusing lens
($f=4.51$\,mm).

If no atom is present in the trap, the second lens re-collimates back the probe
beam. It then passes through several filters and finally is coupled to another
single mode fiber with 72\% efficiency (from B to C in Figure~\ref{exp_setup}).
The other end of the fiber is attached to a silicon avalanche photodiode
(APD2) operating in a passively quenched photon counting mode (dead time about
3\,$\mu$s, quantum efficiency 55\%).
Backscattered light (and atomic fluorescence from the MOT beams) is also
collected into another single mode fiber that is coupled to a second avalanche
photodiode (APD1). Both photodiode signals and a reference signal for the
optical excitation pulses are time-stamped with a resolution below 1\,ns for
further analysis.

Optical excitation pulses with a well-defined envelope are prepared from
a continuous laser locked to the $5S_{1/2}\ket{F=3}
\rightarrow 5P_{3/2}\ket{F=4}$ transition in $^{85}$Rb. An acousto-optical
modulator (AOM) in double-pass configuration ($f\approx200$\,MHz) acts as a
first light switch with an on/off ratio of about 50\,dB, and an electro-optical
modulator (EOM) that generates optical sidebands at 1.5\,GHz away from the
carrier creates the pulses. The pulse shape itself is determined by modulating the radiofrequency
amplitude of the EOM, and can be chosen either as a rising exponential, or an
approximate square profile. The first red optical sideband of the light
leaving the EOM is extracted with a series of three temperature-tuned etalons
(line width $\approx$460\,MHz), suppressing the optical
carrier by about 60\,dB. Details can be found in \cite{expopulse:rsi}.

\begin{figure}
  \includegraphics[width=\columnwidth]{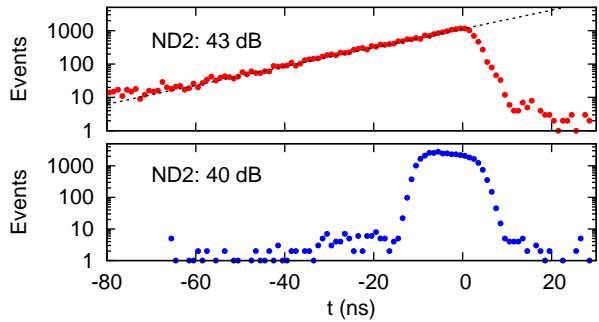}
  \caption{Histograms of detection times with respect to a pulse edge for
    $1.5\cdot10^7$ excitation pulses in time bins of
    1\,ns. Top: exponentially rising pulse with $\tau=15$\,ns and mean
    photon number  $\langle N \rangle=110\pm6$, with an exponential fit
    (dashed line). Bottom: reference pulse with $\tau=15$\,ns and $\langle N
    \rangle=104\pm5$. \label{example_pulses}}
\end{figure}

The average number of photons $\langle N\rangle$ in each pulse was varied by
inserting calibrated neutral density filters (ND1), and determined from a
histogram of detection times of the forward photodiode (see
Fig.~\ref{example_pulses} for two typical pulses). For that, we use the
expression $\langle N\rangle=r_d/(\eta_l\eta_{\mathrm{ND2}})$, where $r_d$ is
the fraction of all pulses that cause photodetection events in 
APD2, $\eta_{\mathrm{ND2}}$ is the transmission of neutral density filter ND2,
and $\eta_l=0.30\pm0.02$ the system efficiency capturing
reflection and coupling losses from A to C (see Fig.~\ref{exp_setup}) and the
quantum efficiency of APD2. To avoid dead time effects of the photodetector,
$\eta_{\mathrm{ND2}}$ was chosen between -25 and -51\,dB such that $r_d
\lesssim 1\%$.

% \section{Experiment}

We start the excitation experiment, once an atom is loaded from the
magneto-optical trap into the FORT, identified by its fluorescence detected
with APD1, with 5\,ms of molasses cooling. Then we optically pump the atom to
$5S_{1/2}\ket{F=2, m_F=-2}$ with a circularly polarized
optical pumping light for
10\,ms. This is followed by a train of 100 optical probe pulses separated by
12\,$\mu$s to minimize dead time effects of the photodetectors.
To keep the central optical frequency  of the pulse
resonant with the $5S_{1/2}\ket{F=2,m_F=-2}\rightarrow
5P_{3/2}\ket{F=3, m_F=-3}$ transition, Zeeman and AC stark shift in
the trap were calibrated in an independent probe transmission measurement with
cw light.

After probing, we verify the presence of the atom by fluorescence from the
molasses beams for 20 to 30\,ms, which also removes the recoil energy the atom
acquired during the probe period. If the atom was not lost, the  probing sequence is
repeated. Otherwise, the same sequence of probe pulses is recorded for 3
seconds as reference to measure the average photon number $\langle N\rangle$
in the pulse.

\section{Results}
The probability $P_e(t)$ of an atom being in the excited state at any time $t$
can be directly assessed by the fluorescence detected in backwards direction
with APD1, because there is no interference between backscattered and
excitation light. We sort the photodetection events into time bins
of width $\Delta t=1$\,ns with respect to the pulse edge. The total number of excitation
pulses $N_T$ that are sent to an atom while it is in the trap is independently
measured by a timestamp unit as shown in Figure~\ref{exp_setup}. The excitation probability in time bin $t$ 
is then given by 
\begin{equation}
\label{eq:exc_prob_vs_rate}
  P_e(t)=N_d(t)/\left(\Gamma_p \Delta t \eta_r N_T \right)\,,
\end{equation}
where $N_d(t)$ is the number of detected fluorescence events in the same time bin,
and $\eta_r=0.30\pm0.02$ is a product of the quantum efficiency of
detector APD1 and the transmission through all optical components from the
atom to the detector. The atomic decay rate into the excitation pulse mode $\Gamma_p$
is proportional to the free space spontaneous decay rate $\Gamma$ of the excited state,
\begin{equation}
\Gamma_p =\eta_p \Gamma,
\end{equation}
$\eta_p$ being the spatial overlap parameter between the atomic dipole and excitation pulses.

Following the analysis of our experimental configuration in~\cite{Tey:09},
the spatial overlap can be expressed in terms of the scattering ratio
$R_{sc}$, which depends on the focusing strength $u:=w_L/f$ as
\begin{equation}
\eta_p=\frac{R_{sc}}{4}=\frac{3}{16u^3}e^{2/u^2}\left[\Gamma\left(-\frac{1}{4},\frac{1}{u^2}\right)+u\Gamma\left(\frac{1}{4},\frac{1}{u^2}\right)\right]^2,
\end{equation}
where $w_L$ is the input beam waist, $f$ the focal distance of the coupling
lens, and $\Gamma(a,b)$ % = \int_b^{\infty} t^{a-1} e^{-t} d t $
the incomplete gamma function.
The parameter $u=0.22$ in our experiment would correspond to $\eta_p = 0.03$.
To verify this number, we recorded the phododetection rate in backward
detector APD1 from a free decay of a single atom excited with a high
probability by a short excitation pulse. This provides a
lower bound $\eta_p=0.027$ for the spatial overlap parameter, which is 
very close to calculated value.

\begin{figure}
  \begin{center}
    \includegraphics[width=\columnwidth]{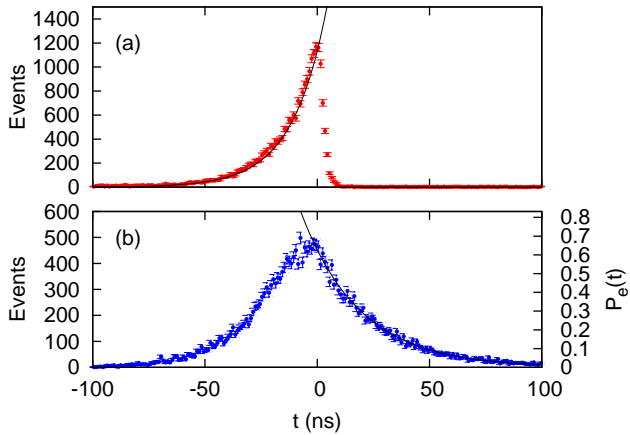}
  \end{center}
  \caption{Exponentially rising excitation pulse (a) and atomic fluorescence
    detected in backward direction (b). The left axis depicts the histogrammed
    photoevents in a 1\,ns wide time bin, the right axis on the fluorescence
    plot the excitation probability derived from it (see text for
    details). Error bars indicate Poissonian counting statistics. The solid
    lines indicate exponential fits with time constants of $\tau=15.4$\,ns for
    the excitation pulse, and $\tau=26.2$\,ns for the decay in the fluorescence.
    \label{fluor_exp}}
\end{figure}

Figure~\ref{fluor_exp}(b) shows the histogram of detection events for an
exponentially rising envelope with a characteristic time $\tau=15$\,ns for
$N_T=2,103,400$ pulses, together with the derived instantaneous excitation
probability $P_e(t)$. As a reference, Fig.~\ref{fluor_exp}(a) shows the
histogram of forward detection events after the atom was lost (with
$\eta_{\mathrm{ND2}}=-43$\,dB) for $N_T\approx1.5\cdot10^7$ % $N_T=14750000$
pulses from which we determined the average photon number $\langle
N\rangle=104.1\pm4.3$ in the excitation pulse as described previously.

During the increasing pulse amplitude, the photoevents in backward
direction also increases exponentially. The atomic population seems to follow the excitation pulse, indicating that we are still in the regime of coherent
scattering for $\langle N \rangle\approx 100$. With this power, we can
transfer $\approx 70\%$ of atomic population to the excited state. After the
excitation field is switched off, the atomic excited state population starts
to decay, leading to an exponentially falling amplitude of light field in
accordance with the Weisskopf-Wigner model \cite{Weisskopf:30}.
We observe atomic fluorescence during the rising excitation
pulse -- in an exact time-reversal of the Weisskopf-Wigner model, however, one
would expect no outgoing field component during that time
\cite{Sondermann:07, Heugel:10}. Due to the limited overlap of the spatial modes
for excitation and emission in our experiment, the destructive interference
between these modes necessary for supression of the scattering is incomplete,
providing an explanation why we still are able to observe fluourescence at that time.

\begin{figure}
  \begin{center}
    \includegraphics[width=\columnwidth]{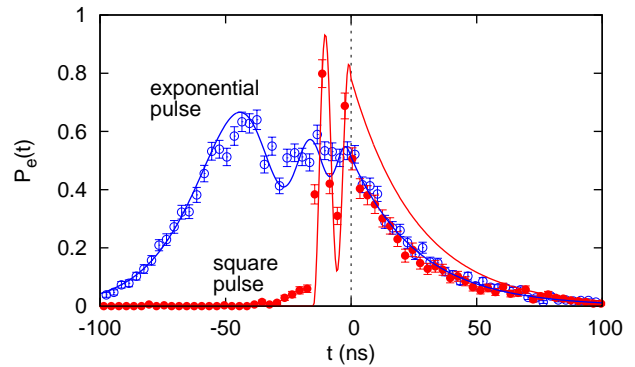}
    \caption{Time-dependent atomic excitation probability $P_e$ for pulses
      with a large average
      photon number $\langle N\rangle\approx1300$ and $\tau=15$\,ns, leading
      to Rabi oscillations. Solid lines represent simulations according
      to the model in \cite{Wang:11}.\label{rabi_flops}}
\end{center}
\end{figure}

With increasing $\langle N\rangle$ in the excitation pulse, the response of
the atom becomes nonlinear, and eventually, the atomic population will undergo
Rabi oscillations. Figure~\ref{rabi_flops} shows such oscillations in atomic
fluorescence both for square and exponential pulses with $\tau=15$\,ns for
$\langle N\rangle\approx1300$.
The oscillation is more pronounced for a square pulse, since during the
long rise time of the exponential pulse, the probability of losing coherence
due to spontaneous emission is larger. The observation of Rabi oscillations
demonstrates a possibility of using a single atom for low photon number
optical switching~\cite{Hwang:09, Englund:11, Volz:11}.

A detailed theoretical analysis of the excitation probability
$P_e(t)$ of a two-level atom by a travelling light pulse in free space can be
found in~\cite{Wang:11,Stobinska:09,CCT:92}. The atomic excitation varies for
different temporal shapes of excitation pulses because
$P_e(t)$ is determined by the dynamical coupling strength
\begin{equation}
g(t)=\sqrt{\Gamma_p\,\langle N\rangle}\,\xi(t)\,.
\label{coupling}
\end{equation}
The normalized temporal envelope functions $\xi(t)$ used to model our
experiments  are
\begin{equation}
   \xi(t) = \left\{
     \begin{array}{ll}
       \frac{1}{\sqrt{\tau}} \exp\left(\frac{1}{2 \,\tau}\,t\right) & \text{for} \,t < 0 \\
    0  & \text{for} \, t > 0
     \end{array}
   \right.
\end{equation}
for the rising exponential, and
\begin{equation}
  \xi(t) = \left\{
    \begin{array}{ll}
    \frac{1}{\sqrt{\tau}}  & \text{for} \,-\tau \leq t \leq 0 \\
    0  & \text{otherwise}
     \end{array}
   \right.
\end{equation}
for square pulse shape. In this model, the other parameter besides
$\langle N\rangle$ determining the coupling strength for a given pulse shape
is the pulse duration $\tau$.

\begin{figure}
  \begin{center}
    \includegraphics[width=\columnwidth]{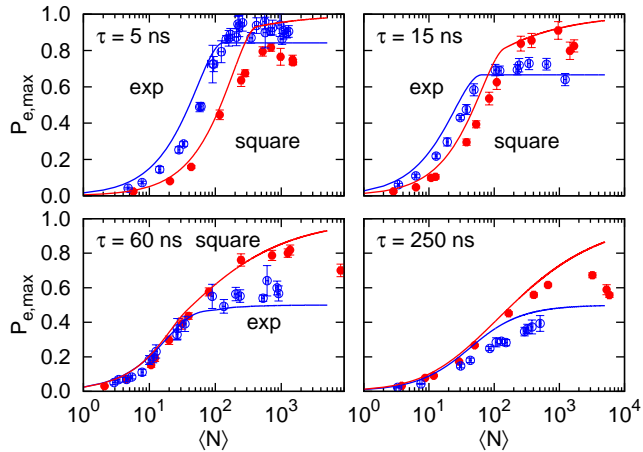}
    \caption{Maximal atomic excitation probability during a single pulse for
      different average photon numbers $\langle N\rangle$ and characteristic
      pulse times $\tau$. Solid lines represent simulations using
      a time-dependent coupling strength of eq.~(\ref{coupling})
      \cite{Wang:11}. \label{sat_curves}}
\end{center}
\end{figure}

To capture the transition between single
excitation and Rabi oscillation over a wide range of parameters,
we consider the maximal excitation probability $P_{e,{\text{max}}}$
during the whole pulse period. A few characteristic traces for both pulse
shapes are shown in Fig.~\ref{sat_curves}.
For long pulses ($\Gamma\tau\gg1$) and with large $\langle N\rangle$,
the atomic population reaches the steady state value of 50\%
expected for a saturated cw excitation. For shorter pulses,
the spontaneous emission probability during the pulse is reduced, and we
observe a higher $P_{e,{\text{max}}}$ for a smaller $\langle N\rangle$, i.e., shorter pulses
are better suited to completely excite the atom. In the regime with low
$\langle N\rangle$, the exponential pulse shape always leads to a higher
excitation probability than the square pulse.

\begin{figure}
  \begin{center}
    \includegraphics[width=\columnwidth]{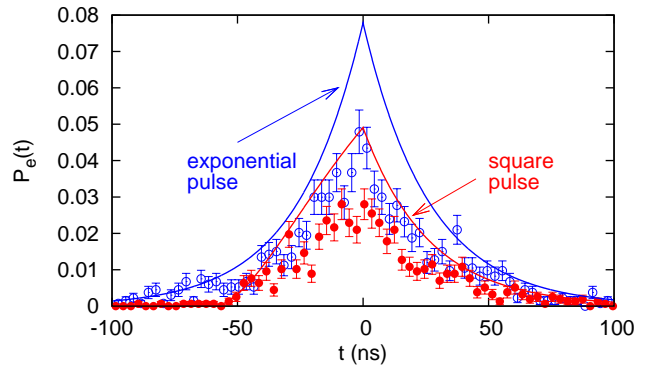}
    \caption{Excitation dynamics for an exponential pulse with
      $\tau_e=25$\,ns and a square pulse with $\tau_{\text{sq}}=60$\,ns
      optimized for the same experimental parameter $\eta=0.027$, and
      comparable $\langle N_e\rangle = 2.75\pm0.06$ and $\langle N_s\rangle =
      2.10\pm0.08$, respectively. Solid lines show theoretical predictions for
      those cases.
     \label{shape_dep}}
\end{center}
\end{figure}

A direct comparison between a square
pulse of width $\tau_s$ and an exponential pulse of rise
time constant $\tau_e=\tau_s$ may not be adequate. We thus compare excitation
probabilities with similar photon numbers, $\langle N_e\rangle =
2.75\pm0.06$ for the exponential, and $\langle N_s\rangle =
2.10\pm0.08$ for the square pulse for $\tau$ that maximize
$P_{e,{\text{max}}}$ for $\eta_p=0.027$. Following~\cite{Wang:11},
maximal excitation would happen for $\tau_e = 24$\,ns and
$\tau_s = 64$\,ns, respectively. The closest available data sets in our
measurements of $P_e(t)$ with  $\tau_e=25$\,ns and $\tau_s=60$\,ns is shown in
Fig.~\ref{shape_dep}. The exponential pulse still leads to larger
$P_{e,{\text{max}}}$ than square pulse for almost the same average photon
number. In these measurements, however, there is a significant difference in
overall amplitude between the model (solid lines) and the measurements, which
we  attribute to residual motion of the atom due to thermal
motion~\cite{colin:10}.

\section{Conclusion}
We have investigated the interaction of temporally shaped pulses with a single
trapped atom, and demonstrated that a single atom can be excited with high
probability using coherent light pulses with relatively low mean photon
number.
The excitation of the atom is sensitive to the envelope of the excitation
pulse, and we experimentally demonstrated that a rising exponential
envelope leads to higher excitation probability in weak excitation regime
which is  compatible with the expectation from a time-reversed
Weisskopf-Wigner process.

According to the theoretical model for the excitation process~\cite{Wang:11},
the advantage of the exponential pulses should become even more prominent for
a larger overlap between excitation and dipole emission modes, as it may be
realized in experiments in~\cite{Sondermann:07, Golla:12}, or for a
replacement of the coherent states with Fock states of the light filed, which
will be the ultimate target in an atom-light interaction for quantum
information processing purposes.

\section{Acknowledgment}
We acknowledge the support of this work by the National Research Foundation \&
Ministry of Education in Singapore.

% \bibliography{../atoms}

\end{document}